# Phase transition in $Pr_{0.5}Ca_{0.5}CoO_3$ and related cobaltites

J. Hejtmánek[1], Z. Jirák[1], O. Kaman[1], K. Knížek[1], E. Šantavá[1], K. Nitta[2], T. Naito[3], and H. Fujishiro[3]

1. Institute of Physics, ASCR, Cukrovarnicka 10, 162 00 Prague 6, Czech Republic, E-mail: hejtman@fzu.cz

2. Japan Synchrotron Radiation Research Institute, Sayo, Hyogo 679-5198, Japan, E-mail: nittak@spring8.or.jp

3. Faculty of Engineering, Iwate University, 4-3-5 Ueda, Morioka 020-8551, Japan, E-mail: fujishiro@iwate-u.ac.jp

## Abstract

We present an extensive investigation (magnetic, electric and thermal measurements and X-ray absorption spectroscopy) of the $Pr_{0.5}Ca_{0.5}CoO_3$ and $(Pr_{1-y}Y_y)_{0.7}Ca_{0.3}CoO_3$ (y = 0.0625 – 0.15) perovskites, in which a peculiar metal-insulator (M-I) transition, accompanied with pronounced structural and magnetic anomalies, occurs at 76 K and 40 – 132 K, respectively. The inspection of the M-I transition using the XANES data of Pr $L_3$-edge and Co $K$-edge proofs the presence of $Pr^{4+}$ ions at low temperatures and indicates simultaneously the intermediate spin to low spin crossover of Co species on lowering the temperature. The study thus definitively confirms the synchronicity of the electron transfer between $Pr^{3+}$ ions and $Co^{3+/4+}O_3$ subsystem and the transition to the low-spin, less electrically conducting phase. The large extent of the transfer is evidenced by the good quantitative agreement of the determined amount of the $Pr^{4+}$ species, obtained either from the temperature dependence of the XANES spectra or via integration of the magnetic entropy change over the $Pr^{4+}$ related Schottky peak in the low-temperature specific heat. These results show that the average valence of $Pr^{3+}/Pr^{4+}$ ions increases (in concomitance with the decrease of the formal Co valence) below $T_{MI}$ for $(Pr_{0.925}Y_{0.075})_{0.7}Ca_{0.3}CoO_3$ up to 3.16+ (the doping level of the $CoO_3$ subsystem decreases from 3.30+ to 3.20+), for $(Pr_{0.85}Y_{0.15})_{0.7}Ca_{0.3}CoO_3$ up to 3.28+ (the decrease of doping level from 3.30+ to 3.13+) and for $Pr_{0.5}Ca_{0.5}CoO_3$ up to 3.46+ (the decrease of doping level from 3.50+ to 3.27+).

## 1. Introduction

The hole doped cobaltites of the general formula $Ln_{1-x}Ae_xCoO_3$ (x>0.2; Ln, Ae – lanthanide and alkali earth ion, respectively) exhibit generally a robust metallic state with ferromagnetic ordering at low temperatures. As an exception, a sharp metal-insulator (M-I) transition at $T_{MI}$ ~ 90 K has been observed on the "half-doped" $Pr_{0.5}Ca_{0.5}CoO_3$ by Tsubouchi et al. [1] and, later on, also on related compounds of lower dopings. It turns out now that the transition is conditioned not only by presence of praseodymium ions, but also by a suitable structural distortion which depends on the average ionic radius and size mismatch of the perovskite $A$-site ions [2]. Furthermore, the stability of the low-temperature insulating phase can be controlled by external means. In particular the high pressure increases $T_{MI}$ and may even induce the transition in some samples with otherwise stable metallic state [3]. On the other hand, $T_{MI}$ is lowered by magnetic field when the "metallic" state can be finally stabilized under a high magnetic field, in particular at 8 Tesla in $(Pr_{1-y}Y_y)_{0.7}Ca_{0.3}CoO_3$ for y = 0.0625 [4]. Similarly, the significant oxygen deficiency in polycrystalline $Pr_{0.5}Ca_{0.5}CoO_{3-\delta}$ samples can lead to complete suppression of the $T_{MI}$ as recently shown by Yoshioka et al. [5]. For such highly oxygen deficient sample the analysis of the Pr-$L_3$ and Co-$K$ X-ray absorption near-edge structure (XANES) complemented by the first-principle calculations revealed, that the valence state of Co ions is instead of 3.5+ (stoichiometic $Pr_{0.5}Ca_{0.5}CoO_3$) only 3.106+ leading to considerable oxygen deficiency of $\delta$~0.197. This observation corroborates both the absence of the metal-insulator transition and the trivalent character of Pr ions, derived from Pr-$L_3$ spectra, and underlines the necessity of the control of the oxygen stoichiometry during the preparation procedure.

The M-I transition observed in the mentioned praseodymium-based cobaltites is further manifested by a pronounced peak in the specific heat, by striking susceptibility drop and large volume contraction. Although these signatures bear resemblance to common spin-state crossover of Co ions, there is a growing evidence that inherent part of the transition is a significant drop of the hole doping, formally from the mixed valence $Co^{3+}/Co^{4+}$ closer to pure $Co^{3+}$, enabled by closeness in energy of the $Pr^{4+}$ and $Pr^{3+}$ states. This scenario was originally suggested in order to account for anomalous shortening of some Pr-O bondlengths, whereas the expected shrink of the $CoO_6$ octahedra due to the stabilization of lower spin states of Co has not been observed [6]. The existence of $Pr \rightarrow CoO_3$ electron transfer was supported later theoretically by GGA+U electronic structure calculations exploiting the temperature dependence of the structural experimental data for $Pr_{0.5}Ca_{0.5}CoO_3$ [7], and experimentally, in particular by observation of Schottky peak in the low temperature specific heat, related to Zeeman splitting of the ground doublet of Kramers ions $Pr^{4+}$ [8]. The direct proof of the mixed $Pr^{3+}$ /$Pr^{4+}$ valence, was provided for $Pr_{0.5}Ca_{0.5}CoO_3$ using the Pr $L_3$ and $M_{4,5}$ edge X-ray absorption spectroscopy (XAS) by Garcia-Munoz *et al.* and Herrero-Martin *et al.* [9, 10], demonstrating also the spin change on cobalt sites by means of the Co $K$ and $L$ edge spectroscopies [11]. Using the measurement of the temperature dependence of the X-ray absorption near-edge structure (XANES) spectra at the Pr $L_3$ edge on $(Pr_{1-y}Y_y)_{0.7}Ca_{0.3}CoO_3$ system, Fujishiro *et al.* observed similar effect, *i.e.* the presence of mixed $Pr^{3+}$/ $Pr^{4+}$ valence at temperatures below the $T_{MI}$ [12].

In this paper we present an extensive investigation of the $Pr_{0.5}Ca_{0.5}CoO_3$ (50% $Co^{4+}$) and $(Pr_{1-y}Y_y)_{0.7}Ca_{0.3}CoO_3$ (30% $Co^{4+}$) systems by means of magnetic, electric, thermal measurements and X-ray absorption spectroscopy. Both the magnetic entropy associated with the low temperature Schottky-like anomaly in specific heat, quantitatively linked with the presence of $Pr^{4+}$ ions, and the analysis of XANES spectra of Pr $L_3$ edge, point to a significant valence shift of Pr below the $T_{MI}$. In $Pr_{0.5}Ca_{0.5}CoO_3$, the essentially trivalent praseodymium ions at 300 K change to a $Pr^{3+}/Pr^{4+}$ mixture of the average valence $Pr^{3.46+}$ at 10 K, which is significantly higher than the value previously deduced using the Pr $L_3$ and $M_{4,5}$ edge spectra by Herrero-Martin *et al.* (from $Pr^{3.0+}$ at 300 K => $Pr^{3.15+}$ at 10 K) [10, 11] and is also higher than praseodymium valence shifts found by us in the less doped $(Pr_{1-y}Y_y)_{0.7}Ca_{0.3}CoO_3$ systems [8, 12]. The simultaneous inspection of the M-I transition using the XANES spectra of Co $K$-edge confirms the change of Co spin state on crossing the M-I transition.

## 2. Experimental

Polycrystalline $(Pr_{1-y}Y_y)_{0.7}Ca_{0.3}CoO_3$ samples were prepared by a solid-state reaction, the detailed sample preparation procedures and chemical characterization proving both the phase purity and oxygen stoichiometry are described elsewhere [8]. The $Pr_{0.5}Ca_{0.5}CoO_3$ sample was, however, prepared using "soft chemistry route" by means of sol-gel procedure. The stoichiometric amounts of $Pr_6O_{11}$ and $CaCO_3$ with chemically determined metal content were dissolved in nitric acid and mixed with volumetric solution of $Co(NO_3)_2$. After removal of the excessive nitric acid by heating and adjustment of pH to 3 by ammonia, the citric acid ($1.5n_{metals}$) and ethylene glycol ($3n_{metals}$) were added. The precursor was formed by gradual evaporation and subjected to slow heating above 250°C. The pulverized material was calcinated in air at 400°C. After being pressed into pellets, it was sintered at 1200°C, followed with a long-term annealing at 800°C in oxygen atmosphere. The ceramic material was found single phase but oxygen deficient, so that an additional annealing was undertaken in high pressure cell (200 atm $O_2$) at 600°C to achieve an oxygen stoichiometric compound.

The X-ray diffraction proved single perovskite phase of orthorhombic Pbnm symmetry in all the prepared samples. The lattice parameters for $Pr_{0.5}Ca_{0.5}CoO_3$ have been refined to a = 5.3384(6) Å, b = 5.3395(7) Å, c = 7.5415(6) Å, V = 214.96(4) Å$^3$. Among the $(Pr_{1-y}Y_y)_{0.7}Ca_{0.3}CoO_3$ samples the parameters a = 5.3472(9) Å, b = 5.3513(8) Å, c = 7.5584(9) Å, V = 216.28(6) Å$^3$ have been obtained in particular for the y=0.10 composition.

The magnetic and specific heat measurements were performed using Quantum Design MPMS and PPMS devices; the experiments at very low temperatures down to 0.4 K were done using the $He^3$ option. The magnetic susceptibility was measured under an applied field of 0.1 T, employing the zero-field and field-cooled regimes during warming and cooling the sample, respectively. Thermal conductivity, thermoelectric power and electrical resistivity were measured using a four-probe method with a parallelepiped sample cut from the sintered pellet. The measurements were done on sample cooling and warming using a close-cycle cryostat down to 3.5 K, the detailed description of the cell including calibration is described elsewhere [13]. For the XANES measurements, a part of the samples were pulverized, mixed with 3N boron nitride (BN) powder with proper molar ratios in order to optimize the absorption and pelletized 6 mm in diameter and 0.5 mm in thickness. The Pr $L_3$-edge and Co $K$-edge XANES spectra were measured at BL01B1 of SPring-8 in Japan under transmission mode with the detectors of ionization chambers and were obtained at various temperatures from 8 to 300 K using a cryocooler [12]. The valence of Pr ions in perovskite samples was supposed to be 3.0+ at 300 K, for the determination of the mixed $Pr^{4+}$ content below $T_{MI}$, a comparative measurement on the oxygen balanced $Pr_6O_{11}$ was made. The XANES spectra were analyzed using the sum of Lorentz functions and one arctangent function representing the "baseline edge" of continuum excitations. For more detailed description see details in ref. [12].

### 3. Results and Discussion

The M-I transitions in our samples are illustrated by peaks in the specific heat in Fig. 1, located at $T_{MI}$ = 76 K for the prototypical compound $Pr_{0.5}Ca_{0.5}CoO_3$ and at $T_{MI}$ = 40 – 132 K for the $(Pr_{1-y}Y_y)_{0.7}Ca_{0.3}CoO_3$ systems with y = 0.0625 – 0.15. The electrical resistivity of the studied samples is presented in Fig 2 together with the data for ferromagnetic "metallic" $Pr_{0.7}Ca_{0.3}CoO_3$ sample. Although the character of conduction can be somewhat "masked" by the polycrystalline form of samples, we clearly identify two distinct behaviors: (i) the real macroscopically "insulating" state with steeply increasing resistivity at low temperatures, represented by the $(Pr_{1-y}Y_y)_{0.7}Ca_{0.3}CoO_3$ sample y=0.15 and (ii) the behavior typical for "bad" or "highly disordered" metals characterized by slowly increasing but finite resistivity at low temperatures, which is represented by the $Pr_{0.7}Ca_{0.3}CoO_3$ compound. Most interestingly, similar finite resistivity is also observed in the low-temperature "insulating" state of prototypical $Pr_{0.5}Ca_{0.5}CoO_3$. The corresponding ground state of the $Pr_{0.5}Ca_{0.5}CoO_3$ sample is thus quite complex, likely linked to the first order transition associated with magnetic disorder and strong electronic correlations [14]. Let us note that our sample and the original one of Tsubouchi *et al.* [1], exhibit both a thermal hysteresis at M-I transition. For the $(Pr_{1-y}Y_y)_{0.7}Ca_{0.3}CoO_3$ samples, however, the transition shows itself as non-hysteretic within the experimental uncertainty, as documented simultaneously by the thermoelectric power data in Fig. 3. Here, the sharp jump of thermopower coefficient below the $T_{MI}$ indicates that the charge carrier concentration is decreased and their itinerancy is strongly inhibited in the low-temperature phase. In contrast to the 30% doping of holes in the Co subsystem at 300 K, the final level can be estimated to hole concentration between 10 – 20% based on a comparison of the low-temperature thermopower with data on the $Pr_{0.9}Ca_{0.1}CoO_3$ and $Pr_{0.8}Ca_{0.2}CoO_3$ compounds, included also in Fig. 3.

It is worth mentioning to the $Pr_{0.5}Ca_{0.5}CoO_3$ samples and their highly hysteretic transition that the electrical resistivity after a cooling run never recovers the anterior value, and further cycling over the transition gradually increases its absolute value. This fact can be explained supposing that this 1-st order transition is accompanied by high elastic constraints of the crystal lattice, which are at the origin of the gradual deterioration of the ceramic sample when cycling over the transition.

The magnetic data of $Pr_{0.5}Ca_{0.5}CoO_3$ and $(Pr_{1-y}Y_y)_{0.7}Ca_{0.3}CoO_3$ samples are confronted in Figs. 4 and 5. The magnetic susceptibility, plotted as $1/\chi$ *vs.* temperature in the Fig. 4 jumps markedly at $T_{MI}$. At high temperatures, after the subtraction of the paramagnetic contribution of the $Pr^{3+}$ ions, the simple analysis enables us to estimate the effective moment of Co species to $\mu_{eff}^2 \sim 10\ \mu_B^2$. This value matches the theoretical value for intermediate-spin (IS) $Co^{3+}/Co^{4+}$ mixture – see also similar results obtained for the $La_{1-x}Sr_xCoO_3$ samples in the range 300 – 600 K by Wu and Leighton [15]. With decreasing temperature, the inverse susceptibility associated with the cobalt subsystem does not decrease linearly, and thus does not follow the Curie-Weiss behaviour with temperature independent parameters. This fact corroborates the hypothesis that Co ions gradually, *i.e.* starting at temperatures well above the $T_{MI}$, change their spin state from the high-temperature alloy with main weight of IS states towards the "novel" ground-state based on mixture of low-spin (LS) species (LS $Co^{3+}$ is diamagnetic while LS $Co^{4+}$ has spin ½ yielding $\mu_{eff}^2$ ~3 $\mu_B^2$), which is finally stabilized below $T_{MI}$ [8].

To shed more light on the magnetic background of the in principle "utmost" non-magnetic ground-state, we present in the Fig.5 the magnetization loops measured at 2 K. The metallic counterpart $Pr_{0.7}Ca_{0.3}CoO_3$ with FM-like transition at $T_C$~55 K exhibits nearly rectangular hysteresis with large coercivity. For $Pr_{0.5}Ca_{0.5}CoO_3$, the hysteresis is preserved though in smaller extent, whereas the magnetization curves for $(Pr_{1-y}Y_y)_{0.7}Ca_{0.3}CoO_3$ are of Brillouin type without perceptible coercivity and remnance. In addition to observed ferro- and/or Brillouin-like magnetism the superposed paraprocess of significant magnitude is observed in all samples. This low-temperature feature is associated with Van Vleck susceptibility of $Pr^{3+}$ ions in the nonmagnetic ground singlet due to crystal field effects. Our experimentally estimated value $\chi_{VV}$ ~ 0.03 – 0.06 $emu.mol^{-1}Oe^{-1}$ per $Pr^{3+}$ matches the absolute value determined by Knizek *et al.* for $Pr_{0.3}CoO_2$ [16]. Turning back to the open hysteresis loop of the $Pr_{0.5}Ca_{0.5}CoO_3$, we note that its occurrence can be related to the fact that due to some composition inhomogeneity and/or oxygen nonstoichiometry in the sample, not all Co species are transformed into the low spin state. A minor phase of supposedly FM character is thus developed below 70 K and interacts with the major glassy FM phase ($T_f$ ~6.5 K)() [17]. This complication is absent in $(Pr_{1-y}Y_y)_{0.7}Ca_{0.3}CoO_3$ samples and, in our opinion, the observed Brillouin-like magnetism cannot be explained without considering a contribution of magnetic $Pr^{4+}$ ions. More detailed analysis exceeds, however, the scope of the presented work and will be published later.

The thermodynamics of the $Pr_{0.5}Ca_{0.5}CoO_3$ and $(Pr_{1-y}Y_y)_{0.7}Ca_{0.3}CoO_3$ systems has been investigated using the specific heat measurements from room temperature down to 0.4 K (see Fig. 1 above). Important information on the nature of the low-temperature insulating phase is, nonetheless, contained in the temperature range 0.4 – 10 K. As we demonstrated previously for the $(Pr_{1-y}Y_y)_{0.7}Ca_{0.3}CoO_3$ sample with y=0.15, the low-temperature specific heat of insulating Pr-based cobaltites, contrary to their metallic counterparts, is accompanied by Schottky peaks which we associate with the ground doublet of Kramers ion $Pr^{4+}$, split by the internal magnetic field existing in the samples [8]. The existence of internal magnetic field acting on Pr sites is, however, somewhat contradicted by apparent absence of long-range magnetic order of the Co and/or Pr sublattice, according to our neutron diffraction patterns taken on the y=0.15 compound down to 0.2 K. Also for $Pr_{0.5}Ca_{0.5}CoO_3$ no observable long-range order has been detected in the study of Baron-Gonzalez *et al.* [18], though the magnetization curve in present Fig.4 suggests an existence of saturated moment of about 0.3 $\mu_B$ per f.u., which is above the resolution limit of powder neutron diffraction. We consider this point as highly interesting and controversial and, consequently, we intend to analyze it in a separate paper.

The low temperature specific heat for $Pr_{0.5}Ca_{0.5}CoO_3$ ($T_{MI}$ = 76 K) and $Pr_{0.63}Y_{0.07}Ca_{0.3}CoO_3$ (the y=0.10 sample, $T_{MI}$ = 93 K), measured at various magnetic fields, is shown in Fig. 6. To separate the $Pr^{4+}$ related Schottky peaks, a background line is modeled using common contributions of specific heat, the hyperfine nuclear $C_p^{hyp} \sim \alpha T^{-2}$, lattice $C_p^{latt} \sim \beta T^3$ and linear "glassy" $C_p^{glass} \sim \gamma T$ terms, and includes as well a Schottky-like contribution due to thermal excitation of $Pr^{3+}$ ions from the ground singlet to excited one at $\Delta E = 6$ meV [19]. This latter contribution becomes important for T>15 K and peaks at about 30 K. The values actually used are $\alpha \sim 0.060$ $Jmol^{-1}K$, $\beta \sim 0.000130$ $Jmol^{-1}K^{-4}$, $\gamma \sim 0.050$ $Jmol^{-1}K^{-2}$ for $Pr_{0.5}Ca_{0.5}CoO_3$ and $\alpha \sim 0.028$ $Jmol^{-1}K$, $\beta \sim 0.000175$ $Jmol^{-1}K^{-4}$, $\gamma \sim 0.028$ $Jmol^{-1}K^{-2}$ for $Pr_{0.63}Y_{0.07}Ca_{0.3}CoO_3$. The hyperfine nuclear contribution, emphasized in Fig.6 by use of *log-log* scale, is especially large for $Pr_{0.5}Ca_{0.5}CoO_3$. It is characterized by coefficient $\alpha \sim 0.060$ $Jmol^{-1}K$, which is apparently field independent and surpassing by more then one order the nuclear specific heat originating from the spin $I=7/2$ multiplet of $^{59}Co$ nuclei in the ferromagnetic cobaltites [20]. The origin of such anomalously large value may be associated to the contribution of $^{141}Pr$ nuclei with spin $I=5/2$ in the hyperfine field of $Pr^{4+}$ electronic pseudospins [8].

After subtraction of the background, the $Pr^{4+}$ contribution for $Pr_{0.5}Ca_{0.5}CoO_3$ and $Pr_{0.63}Y_{0.07}Ca_{0.3}CoO_3$ samples is plotted as $C_p/T$ *vs.* $T$ in Fig.7. The analysis is made supposing anisotropic $g$-factor of axial symmetry, so that it is described by two components $g_{\|}$ and $g_{\perp}$ only. This model leads to a modified Schottky form, where the energy splitting $\Delta E$ for a particular direction is given by the angle $\theta$ corresponding to the deviation from the direction of the magnetic field. The partial contribution to the overall Schottky-like anomaly is calculated as $[(\Delta E_{\|}\cos\theta)^2 + (\Delta E_{\perp}\sin\theta)^2]^{1/2}$ and the contribution to specific heat is weighted by $\sin\theta$, which corresponds to the random orientation of crystallites in the sample. The fit for $Pr_{0.5}Ca_{0.5}CoO_3$, represented by solid lines in upper panel of Fig.7, gives the $g$ factors for $Pr^{4+}$ species as $g_{\perp} \cong 1.4$ and $g_{\|} \cong 4.0$, respectively. Furthermore the magnetic field dependence of the characteristic energies $\Delta E_{\perp}$ ($\Delta E_{\|}$) = f(B) in Fig. 8 enables us to determine the molecular field experienced by rare earth moments as the intercept with B axis to $B_{mol}(Pr^{4+}) \sim 7.5$ T. For $Pr_{0.63}Y_{0.07}Ca_{0.3}CoO_3$, somewhat larger $g_{\perp} \cong 1.65$ and $g_{\|} \cong 4.7$ are obtained, but the molecular field is smaller, $B_{mol}(Pr^{4+}) \sim 2.5$ T. The most important result of this analysis is, however, offered by the integration of the Schottky specific heat as $\Delta\Sigma^{Schottky} = \int C_p^{Schottky} / T \, dT$, which provides the quantitative determination of entropy change associated with the temperature induced redistribution of $Pr^{4+}$ pseudospins between the splitted levels of the ground doublet. Comparing the evaluated entropy change $\Delta\Sigma^{Schottky} \cong 1.50$ $Jmol^{-1}K^{-1}$ (identical value calculated for all magnetic fields) for $Pr_{0.5}Ca_{0.5}CoO_3$ and $\Delta\Sigma^{Schottky} \cong 0.74$ $Jmol^{-1}K^{-1}$ for $Pr_{0.63}Y_{0.07}Ca_{0.3}CoO_3$ with the theoretical limit $Rln2 = 5.76$ $Jmol(Pr^{4+})^{-1}K^{-1}$, we can estimate the absolute amount of $Pr^{4+}$. This is evaluated as 0.26 per f.u. in the case of $Pr_{0.5}Ca_{0.5}CoO_3$, which gives the "real" chemical formula below the M-I transition as $Pr^{3+}_{0.24}Pr^{4+}_{0.26}Ca^{2+}_{0.5}Co^{3+}_{0.76}Co^{4+}_{0.24}O_3$. Similarly the number of $Pr^{4+}$ ions in the y = 0.10 sample is determined to 0.13 per f.u., corresponding to formula $Pr^{3+}_{0.60}Pr^{4+}_{0.13}Ca^{2+}_{0.3}Co^{3+}_{0.83}Co^{4+}_{0.17}O_3$. This latter value can be compared with previously reported $Pr^{4+}$ contents in the low-temperature phase of $(Pr_{1-y}Y_y)_{0.7}Ca_{0.3}CoO_3$ samples y = 0.075 and 0.15, which amounted to 0.12 per f.u. and 0.18 per f.u., respectively [8].

Apart of this "indirect" proof of the electronic transfer between the $Pr^{3+}$ and $Co^{4+}$ species we have probed directly the evolution of the Pr and Co valency using the XANES spectra of Pr $L_3$- edge and Co $K$ edge. In Fig.9 we present the temperature dependence of the XANES spectra at the Pr $L_3$- edge for the $Pr_{0.5}Ca_{0.5}CoO_3$ sample. The detailed analysis including the fitting procedure goes beyond this work and we refer to the recent paper of Fujishiro *et al.* [12]. Nonetheless it remains clear that the shape of the XANES spectra changes markedly when the temperature crosses the transition temperature $T_{MI}$ = 76 K – see the lower panel of Fig.9. The sudden decrease of the peak situated at 5966 eV on expense of that at 5979 eV indicates unequivocally the appearance of $Pr^{4+}$ species. The quantitative analysis based on the calibration using $Pr_6O_{11}$ standard, Lorentzian deconvolution and arctangent background subtraction enables to trace the temperature evolution of the valence of Pr ions in $Pr_{0.5}Ca_{0.5}CoO_3$, which is shown in comparison with two previously studied $(Pr_{1-y}Y_y)_{0.7}Ca_{0.3}CoO_3$ compounds in Fig.10. Confronting the low temperature chemical formula, derived on a base of XANES results, $Pr^{3+}_{0.3}Pr^{4+}_{0.2}Ca^{2+}_{0.5}Co^{3+}_{0.7}Co^{4+}_{0.3}O_3$, with that based on the Schottky analysis, we find a reasonable agreement. Consequently we estimate the Co valency below the $T_{MI}$ transition supposing the charge neutrality and full oxygen stoichiometry as average between the Schottky and XANES based analysis of Pr valency as $Co^{3.27+(\pm 0.03)}$. This result presumes the electron transfer of ~ 0.23 $e$ between Pr and Co in case of $Pr_{0.5}Ca_{0.5}CoO_3$ which is still higher than the electron transfer of ~ 0.10 and 0.17 $e$ observed for $(Pr_{1-y}Y_y)_{0.7}Ca_{0.3}CoO_3$ systems with y=0.075 and 0.15, respectively.

The spin-state change that occurs concomitantly with the M-I transition has been investigated using XANES at Co $K$ edge. The spectra seen in Fig.11 refer to the $Pr_{0.63}Y_{0.07}Ca_{0.3}CoO_3$ (or y=0.10) sample where the electron transfer of ~ 0.13 $e$ between Pr and Co is determined based on the Schottky peak analysis. Two dominant features can be noticed in the XANES region - the $1s \rightarrow 4p$ main absorption peak at 7726 eV and the pre-edge peak centered around 7710 eV that reflects $1s \rightarrow 3d$ transitions via the p-d orbital mixing. As evidenced in Fig.11, the main absorption peak does not show any detectable shift with decreasing temperature. This is quite surprising considering the significant population of $Pr^{4+}$ ions below the $T_{MI}$ that should be compensated by an appropriate decrease of Co valence. A plausible explanation why the change of Co valence is not directly reflected by the chemical shift of the main peak could be found in an interplay of competing effects between: (i) the chemical shift, which may make about 0.4 eV to low energy side in the case of Co valence change 3.3+ to 3.13+ encountered in $Pr_{0.63}Y_{0.07}Ca_{0.3}CoO_3$, and (ii) the lowering of spin state, which increases the main edge energy for the same valence of Co. The impact of the spin state is difficult estimate quantitatively, but very roughly, based on the paper of Vanko *et al.* [21] on $LaCoO_3$ the shift from LS to HS leads to increase of ~ 0.5 eV.

The pre-edge peak, shown in more detail in the inset of Fig.11, exhibits also little change with decreasing temperature. Compared to the spectra of $LnCoO_3$, that allow a decomposition to $t_{2g}$ and $e_g$ components and the temperature induced spin states can be readily probed [21,22], the pre-edge peak in mixed-valent cobaltites is much broader and rather featureless. The same vague shape applies also for the pre-edge peak of present $Pr_{0.63}Y_{0.07}Ca_{0.3}CoO_3$ sample. The close inspection of the pre-edge feature after the subtraction of the background (shown by the dashed line) reveals, nevertheless, a small transfer of thespectral weight from the low to high energy side at low temperatures, as expected for filling of $t_{2g}$ levels upon the transition to lower spin states. The surplus of $e_g$ component seems to be less diffusive, forming a rather separated bump at 7711.5 eV. We relate this feature, based on the character of pre-peak on the insulating compound $LiCoO_2$ [23], to a partial stabilization of localized LS $Co^{3+}$ states in the $Pr_{0.63}Y_{0.07}Ca_{0.3}CoO_3$ sample. Let us note that practically identical evolution of the Co *K*-edge pre-edge peak was presented most recently for $Pr_{0.5}Ca_{0.5}CoO_3$ by Herrero-Martin *et al.* [11].

## 4. Conclusions

Magnetic, electric, thermal and X-ray absorption spectroscopy data have been accumulated for the $Pr_{0.5}Ca_{0.5}CoO_3$ and $(Pr_{1-y}Y_y)_{0.7}Ca_{0.3}CoO_3$ (y = 0.0625 – 0.15) perovskites, which exhibit the unusual M-I transition at 76 K and between 40 – 132 K, respectively. The study provides conclusive arguments that the transition is accompanied with large electron transfer from the $Pr^{3+}$ ions to the $CoO_3$ subsystem on cooling through the transition. The latter process is quantified by two independent experimental means: (i) by the analysis of Schottky peak in the low temperature specific heat, which are related to Zeeman splitting of the ground-state doublet of Kramers ions $Pr^{4+}$, and (ii) by the comparative XANES investigation at Pr $L_3$-edge with use of the $Pr^{3+}/Pr^{4+}$ standard. The mutually consistent data on the charge transfer between Pr species and $CoO_3$ subsystem are obtained. In particular, for the prototypical compound $Pr_{0.5}Ca_{0.5}CoO_3$, the amount of $Pr^{4+}$ ions at 8 K is determined to 0.23±0.03 per f. u., corresponding to the change of average praseodymium valence from common value 3+ at room temperature to about 3.46+ in the low-temperature phase. This is a surprisingly large valence shift in view of recent report, where a combined analysis of X-ray spectroscopic data, including the behavior at both the Pr and Co absorption or emission edges, gave three times lower change [11]. For $(Pr_{1-y}Y_y)_{0.7}Ca_{0.3}CoO_3$, the amount of $Pr^{4+}$ ions is smaller, but still important: ~ 0.13 per f.u. for the $Pr_{0.63}Y_{0.07}Ca_{0.3}CoO_3$ (y = 0.10) compound.

Somewhat controversial result is obtained in the temperature dependence of XANES spectra at Co K-edge on $Pr_{0.63}Y_{0.07}Ca_{0.3}CoO_3$. Two observations are noticeable. The first one is the apparent lack of chemical shift on cooling the sample below the $T_{MI}$, which is naturally anticipated as a consequence of the electron transfer from Pr species to $CoO_3$ subsystem. This effect might originate eventually from competing tendencies acting oppositely on the position of main edge, i.e the spin state transition and the charge transfer. More important finding is, however, the empirically deduced very small $t_{2g} \rightarrow e_g$ transfer of spectral weight in the pre-edge peak. This observation questions in principal not only the existence of significant shift of Co valence, but also the complete IS → LS crossover, considered as the basic constituent of the M-I transition in Pr-based, mixed valence cobaltites. On the other hand, we are firmly convinced that the electron transfer from Pr sites to $CoO_3$ subsystem does exist, and is manifested in the electric transport properties as significant drop of the hole carrier concentration. In particular for $Pr_{0.63}Y_{0.07}Ca_{0.3}CoO_3$ the observed amount 0.13 $Pr^{4+}$ per. f. u. means that the doping level is decreased from the 30% doping level at room temperature to about 17% doping level below the M-I transition. We thus conclude that the nature of the low-temperature phase is not a simple issue, leaving the question on the carrier character and spin states, as well on the character of $Pr^{4+}$ pseudospins and origin of strong molecular field acting on them, open and worth to be further studied.

**Acknowledgements**

We thank to Prof. Nan Lin Wang of Institute of Physics, Beijing for fruitful discussion and possibility of the high-pressure oxygenation. The work was performed under the financial support of the Grant Agency of the Czech Republic within the Project No.204/11/0713. The synchrotron radiation experiments were performed at the BL01B1 of SPring-8 with approval of Japan Synchrotron Radiation Research Institute (Proposal No. 2011A1060).

**References**

1) S. Tsubouchi, T. Kyômen, M. Itoh, P. Ganguly, M. Oguni, Y. Shimojo, Y. Morii, and Y. Ishii, Phys. Rev. B, **66**, 052418 (2002)
2) T. Naito, H. Sasaki, and H. Fujishiro, J. Phys. Soc. Jpn., **79**, 034710 (2010)
3) T. Fujita, T. Miyashita, Y. Yasui, Y. Kobayashi, M. Sato, E. Nishibori, M. Sakata, Y. Shimojo, N. Igawa, Y. Ishii, K. Kakurai, T. Adachi, Y. Ohishi, and M. Takata, J. Phys. Soc. Jpn. **73,** 1987 (2004)
4) T. Naito, private communication, unpublished
5) T. Yoshioka, T. Yamamoto; and A. Kitada, Jap. J. Appl. Phys., **51,** 073201 (2012)
6) A. J. Baron-Gonzalez, C. Frontera, J. L. Garcıa-Munoz, J. Blasco, and C. Ritter, Phys. Rev. B **81**, 054427 (2010).
7) K. Knížek, J.Hejtmánek, P. Novák, and Z. Jirák, Phys. Rev. B , **81** ,155113 (2010)
8) J. Hejtmánek, E. Šantavá, K. Knížek, M. Maryško, Z. Jirák, T. Naito, H. Sasaki, and H. Fujishiro, Phys. Rev. B, **82**, 165107 (2010)
9) J. L. Garcia-Munoz, C. Frontera, A. J. B. Gonzalez, S. Valencia, J. Blasco, R. Feyerherm, E. Dudzik, R. Abrudan, and F. Radu, Phys. Rev. B **84**, 045104 (2011)
10) J. Herrero-Martín, J. L. Garcıa-Munoz, S. Valencia, C. Frontera, J. Blasco, A. J. Baron-Gonzalez, G. Subıas, R. Abrudan, F. Radu, E. Dudzik, and R. Feyerherm, Phys. Rev. B **84**, 115131 (2011).
11) J. Herrero-Martın, J. L. Garcıa-Munoz, K. Kvashnina, E. Gallo, G. Subıas, J. A. Alonso, and A. J. Baron-Gonzalez, Phys. Rev. B **86**, 125106 (2012)
12) H. Fujishiro, T. Naito, S. Ogawa, N. Yoshida, K. Nitta, J. Hejtmanek, K. Knizek, and Z. Jirak, Journal of the Physical Society of Japan, **81**, 064709 (2012)
13) J. Hejtmánek, Z. Jirák, M. Maryško, C. Martin, A. Maignan, M. Hervieu, and B. Raveau, Phys. Rev. B **60**, 14057 (1999)
14) E Miranda and V Dobrosavljevic, Rep. Prog. Phys. **68**, 2337 (2005)
15) J. Wu and C. Leighton, Phys. Rev. B **67**, 174408 (2003)
16) K. Knížek, Z. Jirák, J. Hejtmánek, M. Maryško, and J. Buršík, J.Appl.Phys., **111**, 07D707 (2012)

17) M. Maryško, Z. Jirák, J. Hejtmánek, and K. Knížek, J.Appl. Phys. **111**, 07E110 (2012)

18) A.J. Barón-González, C. Frontera, J.L. García-Muñoz, J. Blasco, C. Ritter, S. Valencia, R. Feyerherm, and E. Dudzik, Physics Procedia **8**, 73 (2010)

19) A. Podlesnyak, S. Rosenkranz, F. Fauth, W. Marti, H.J. Scheel, and A. Furrer, J. Phys.-Condens. Matter **6**, 4099 (1994)

20) C. He, S. Eisenberg, C. Jan, H. Zheng, J. F. Mitchell, and C. Leighton, Phys. Rev. B **80**, 214411 (2009)

21) G. Vankó, J.-P. Rueff, A. Mattila, Z. Németh, and A. Shukla, Phys. Rev. B **73**, 024424 (2006)

22) O. Haas, R.P.W.J. Struis, and J.M. McBreen, J. Solid State Chem **177**, 1000 (2004)

23) W.-S. Yoon, K.-K. Lee, and K.-B. Kim, J. of Power Sources **97-98**, 303 (2001)

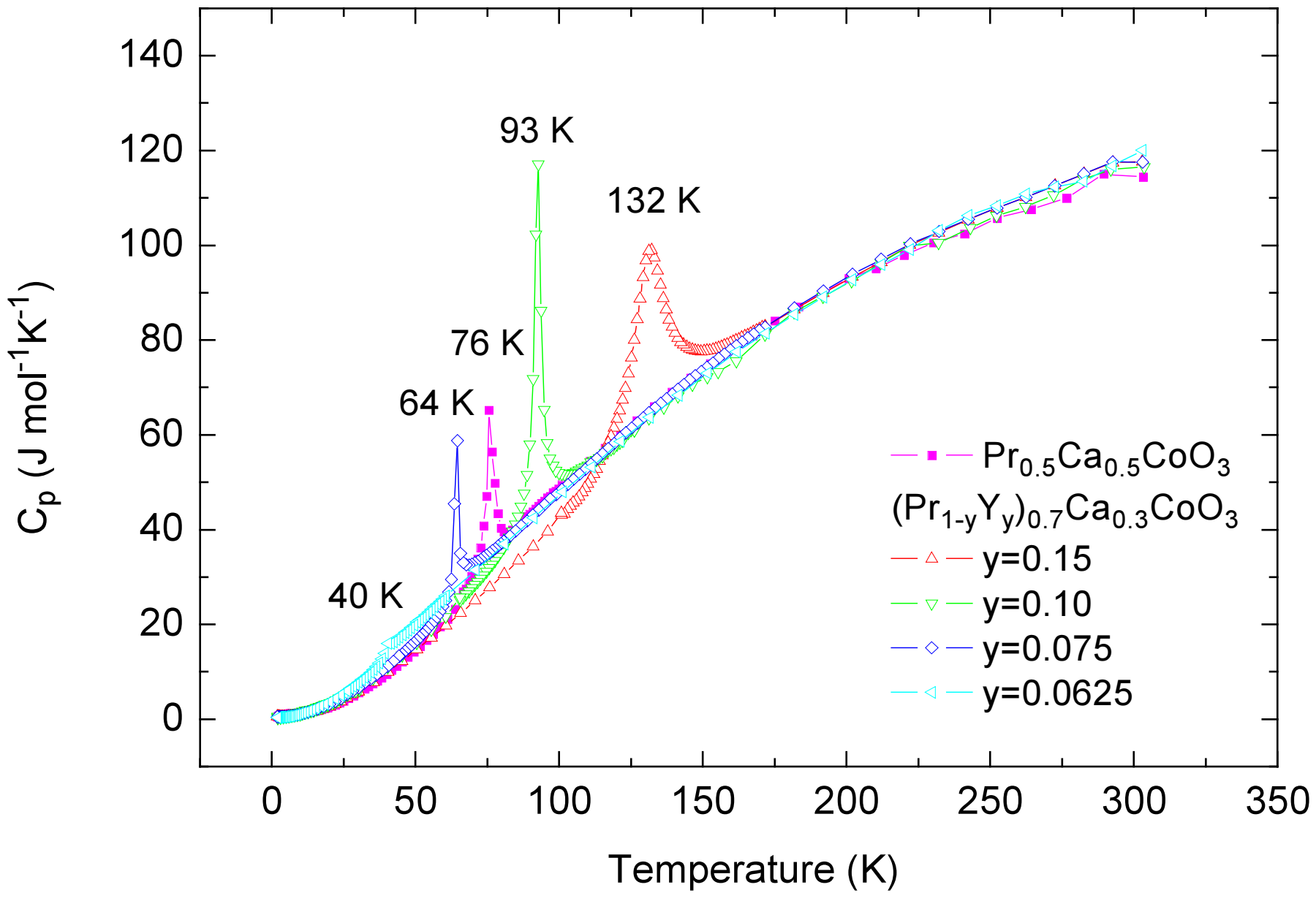

Fig. 1. (Color online) Temperature dependence of the specific heat of $(Pr_{1-y}Y_y)_{0.7}Ca_{0.3}CoO_3$ (y = 0.0625-0.15) and $Pr_{0.5}Ca_{0.5}CoO_3$. The critical temperatures of M-I transitions are marked.

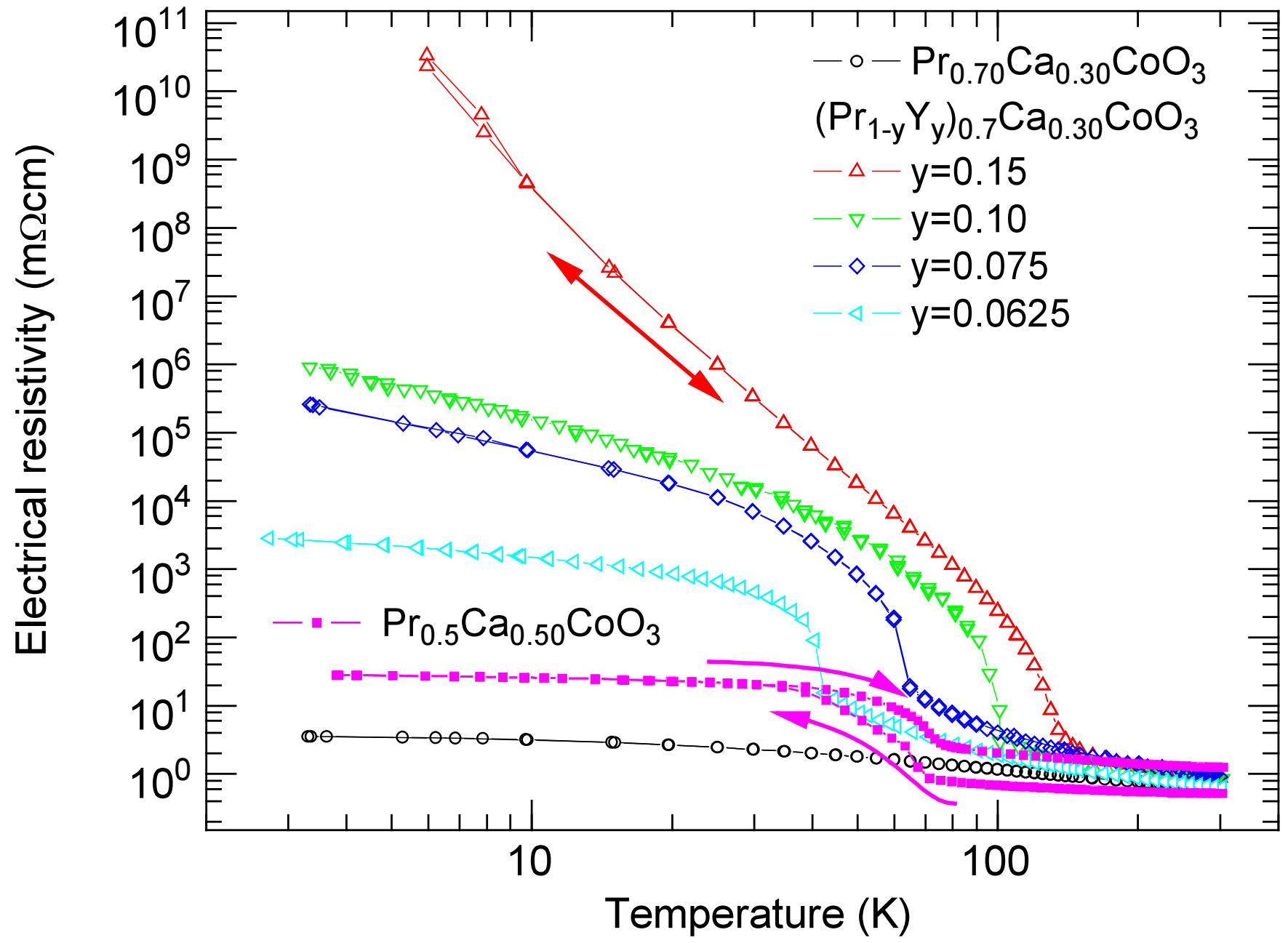

Fig.2. (Color online) The temperature dependence of electrical resistivity of $(Pr_{1-y}Y_y)_{0.7}Ca_{0.3}CoO_3$ (y = 0.0625-0.15) and $Pr_{0.5}Ca_{0.5}CoO_3$, plotted in *log*-*log* scale. The data for $Pr_{0.7}Ca_{0.3}CoO_3$ with ferromagnetic metallic ground state are added for comparison.

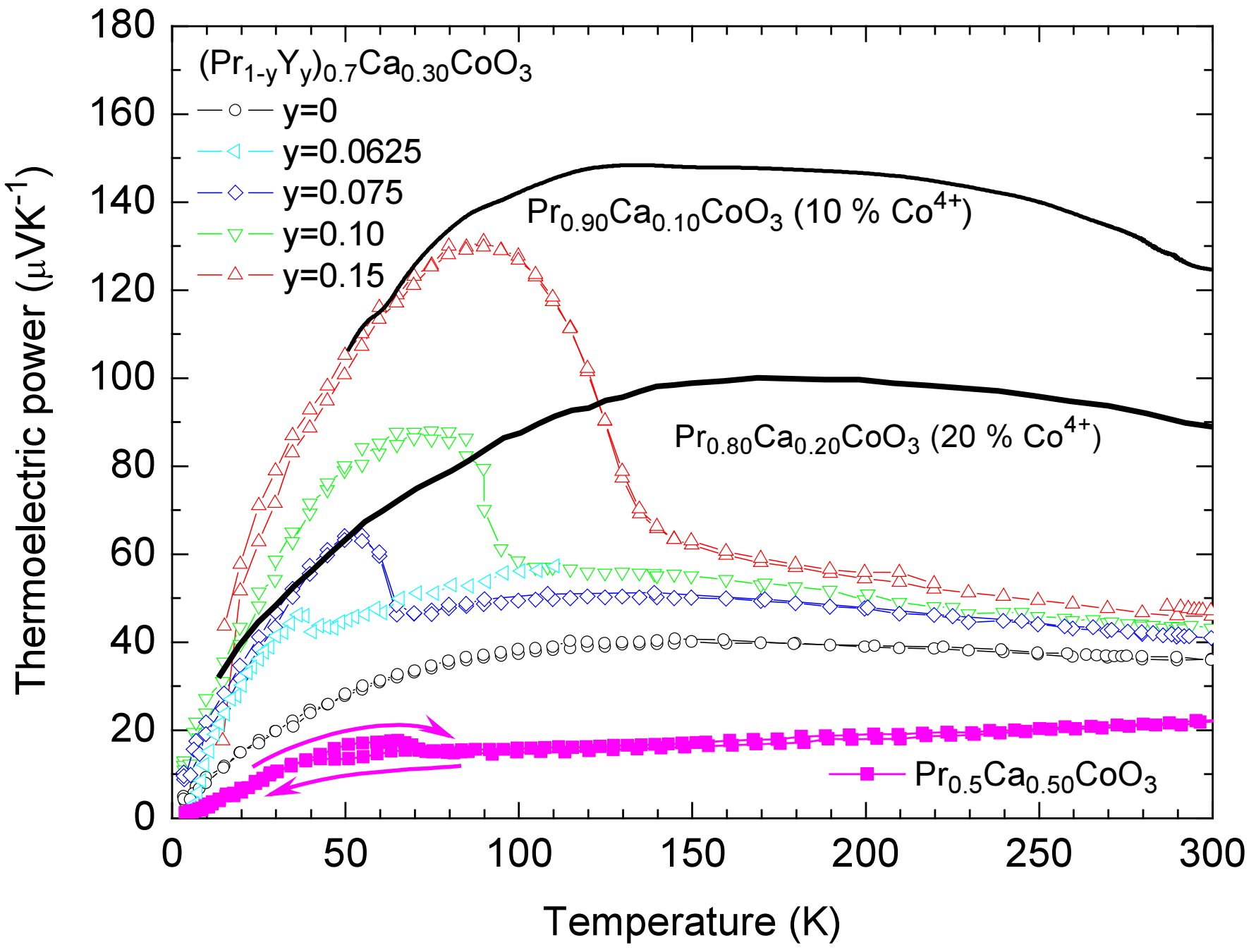

Fig.3. (Color online) The thermoelectric power of $(Pr_{1-y}Y_y)_{0.7}Ca_{0.3}CoO_3$ (y = 0.0-0.15) and $Pr_{0.5}Ca_{0.5}CoO_3$. The transition is evident from the increase below the $T_{MI}$, the data for cobaltites $Pr_{1-x}Ca_xCoO_3$ of lower hole doping are given for comparison.

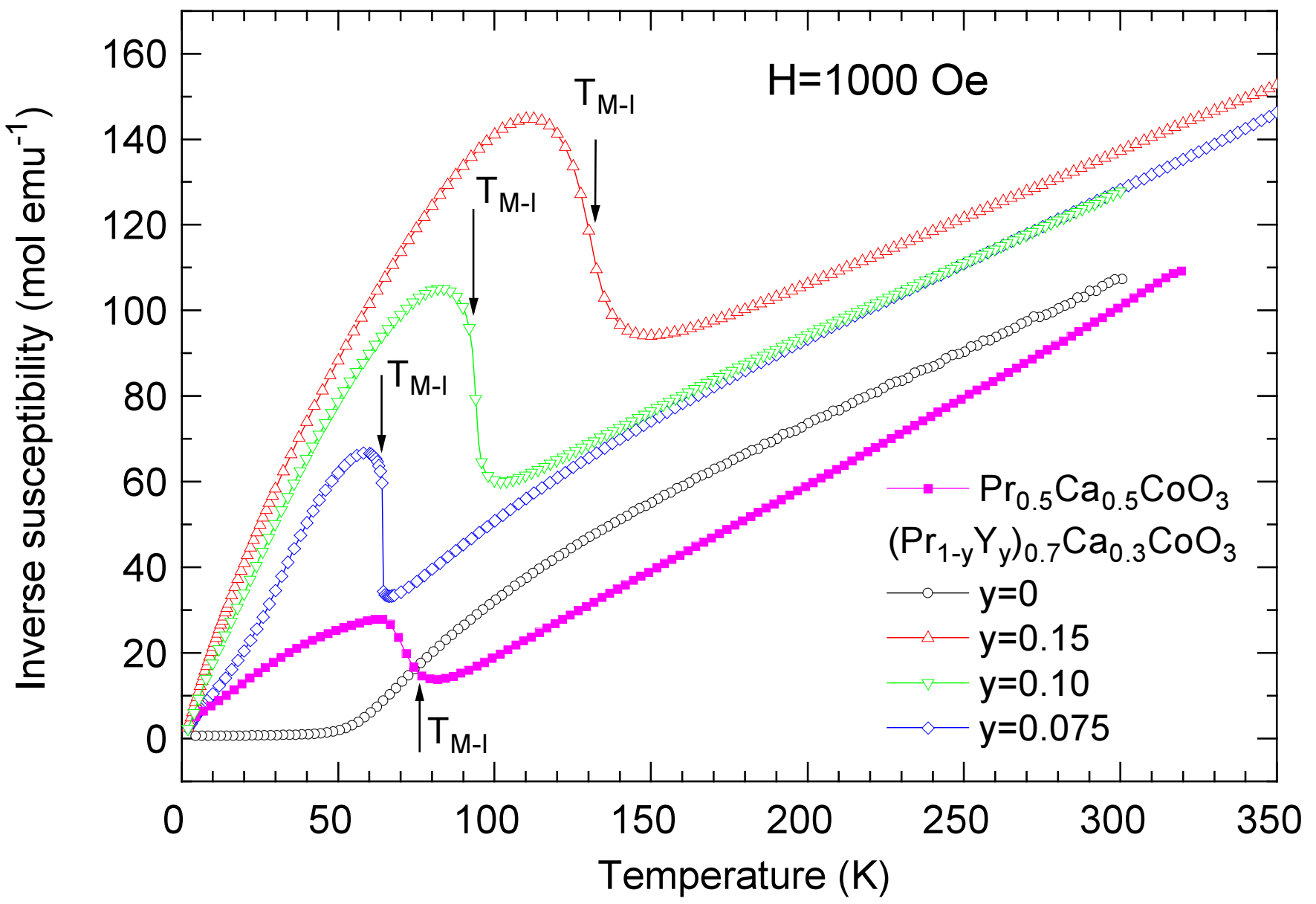

Fig. 4. (Color online) The inverse of the magnetic susceptibility as a function of the temperature for $(Pr_{1-y}Y_y)_{0.7}Ca_{0.3}CoO_3$ (y = 0.0-0.15) and $Pr_{0.5}Ca_{0.5}CoO_3$.

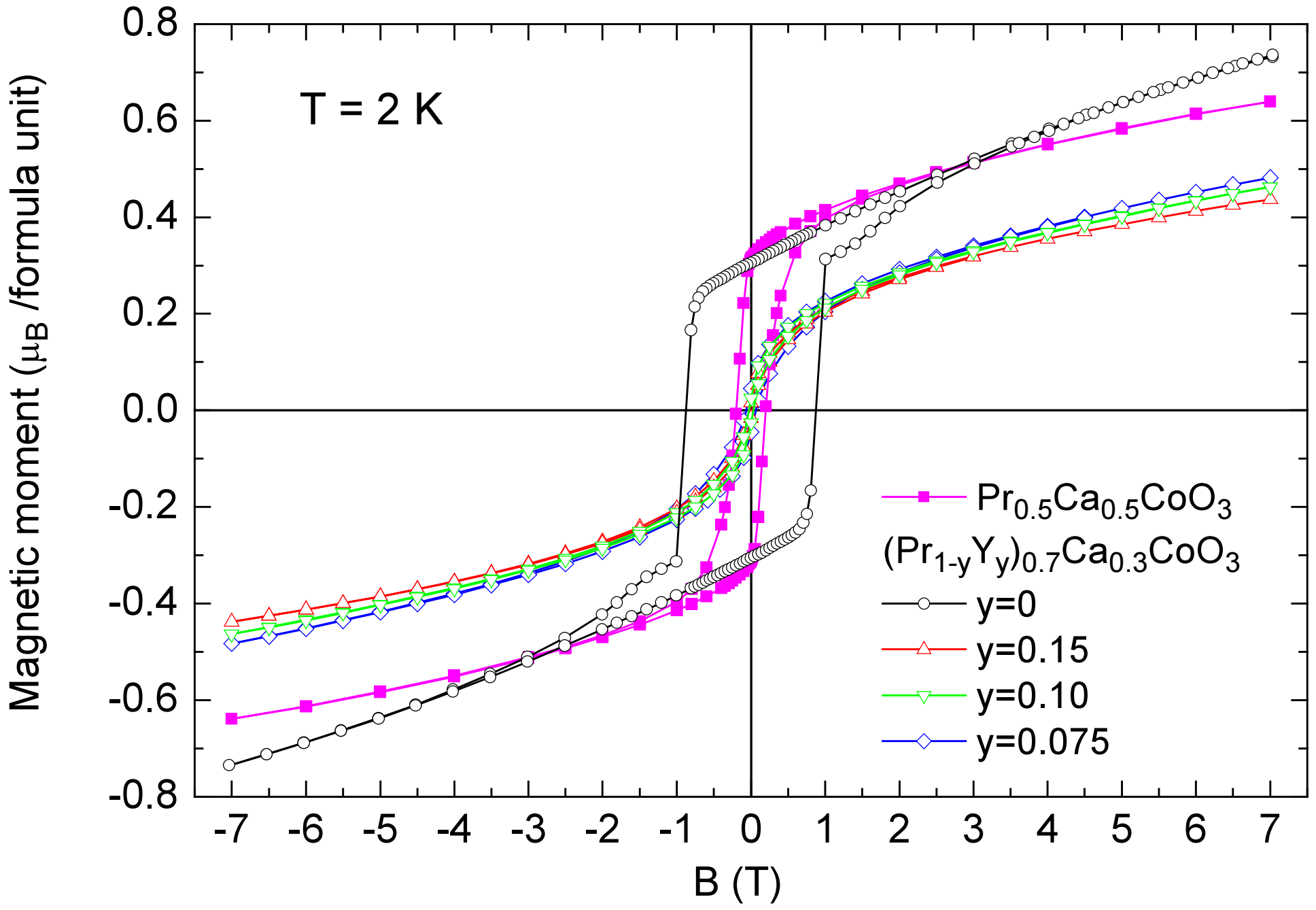

Fig. 5. (Color online) Magnetization loops measured at 2 K for $(Pr_{1-y}Y_y)_{0.7}Ca_{0.3}CoO_3$ (y = 0.0-0.15) and $Pr_{0.5}Ca_{0.5}CoO_3$.

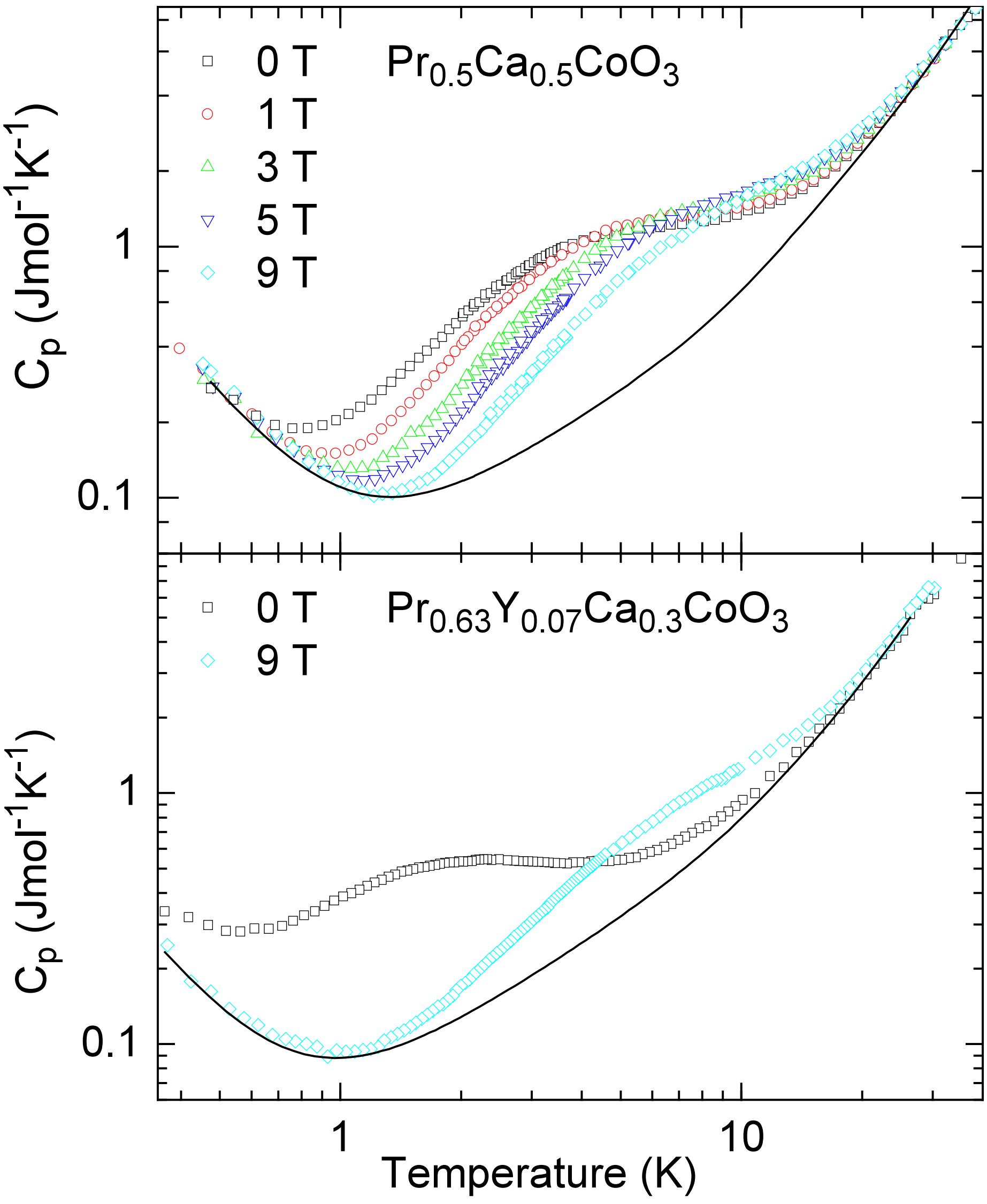

Fig.6. (Color online) The low-temperature specific heat of $Pr_{0.5}Ca_{0.5}CoO_3$ and $Pr_{0.63}Y_{0.07}Ca_{0.3}CoO_3$ measured in fields 0–9 T and plotted in *log*-*log* scale. In order to separate the $Pr^{4+}$ related Schottky peaks, the background contribution comprising common terms is marked by full line (see the text).

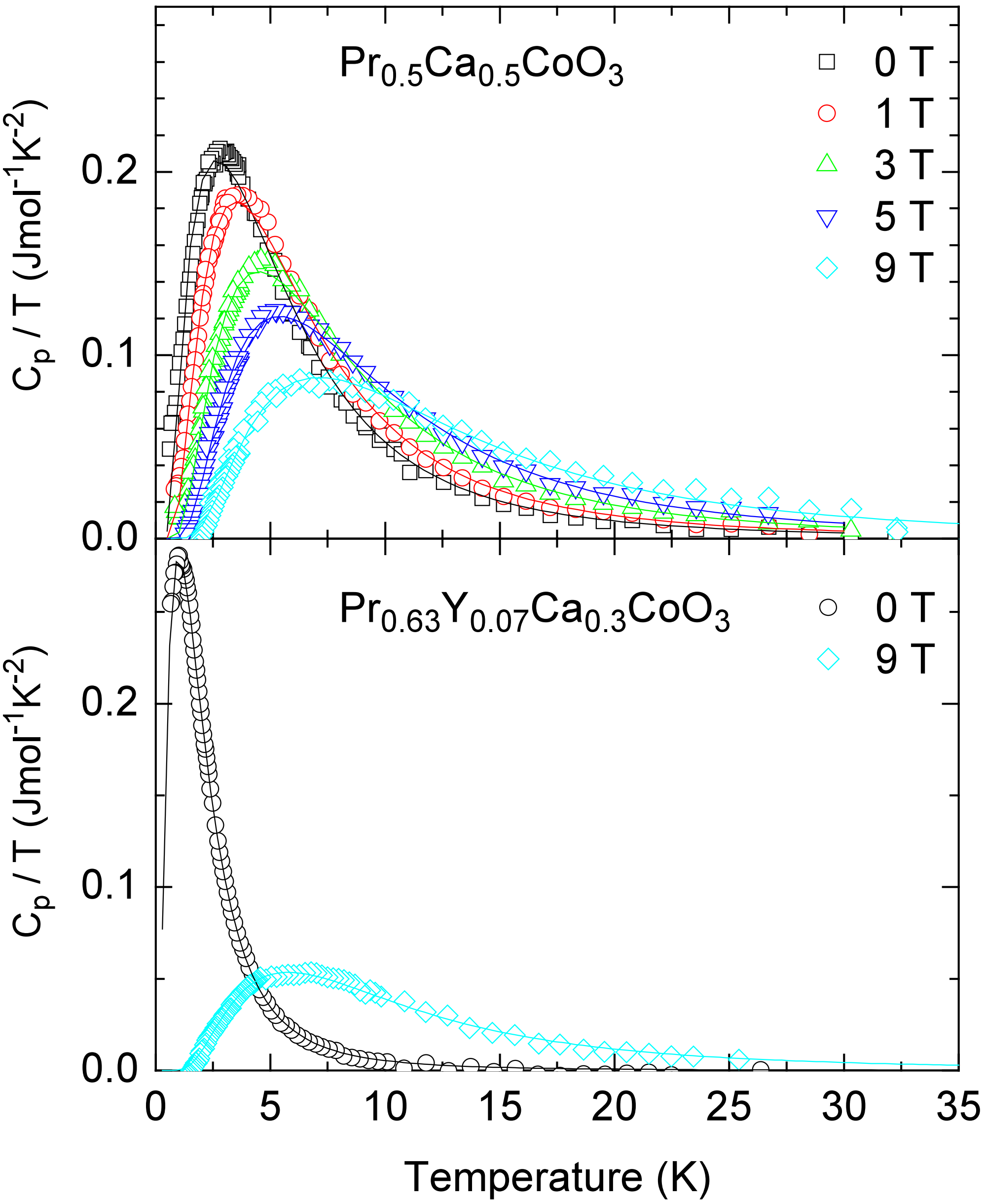

Fig.7. (Color online) The Schottky $Pr^{4+}$ contribution for $Pr_{0.5}Ca_{0.5}CoO_3$ and $Pr_{0.63}Y_{0.07}Ca_{0.3}CoO_3$ plotted as $C_p/T$ vs. T together with the fit based on anisotropic *g*-factor as described in the text (solid lines).

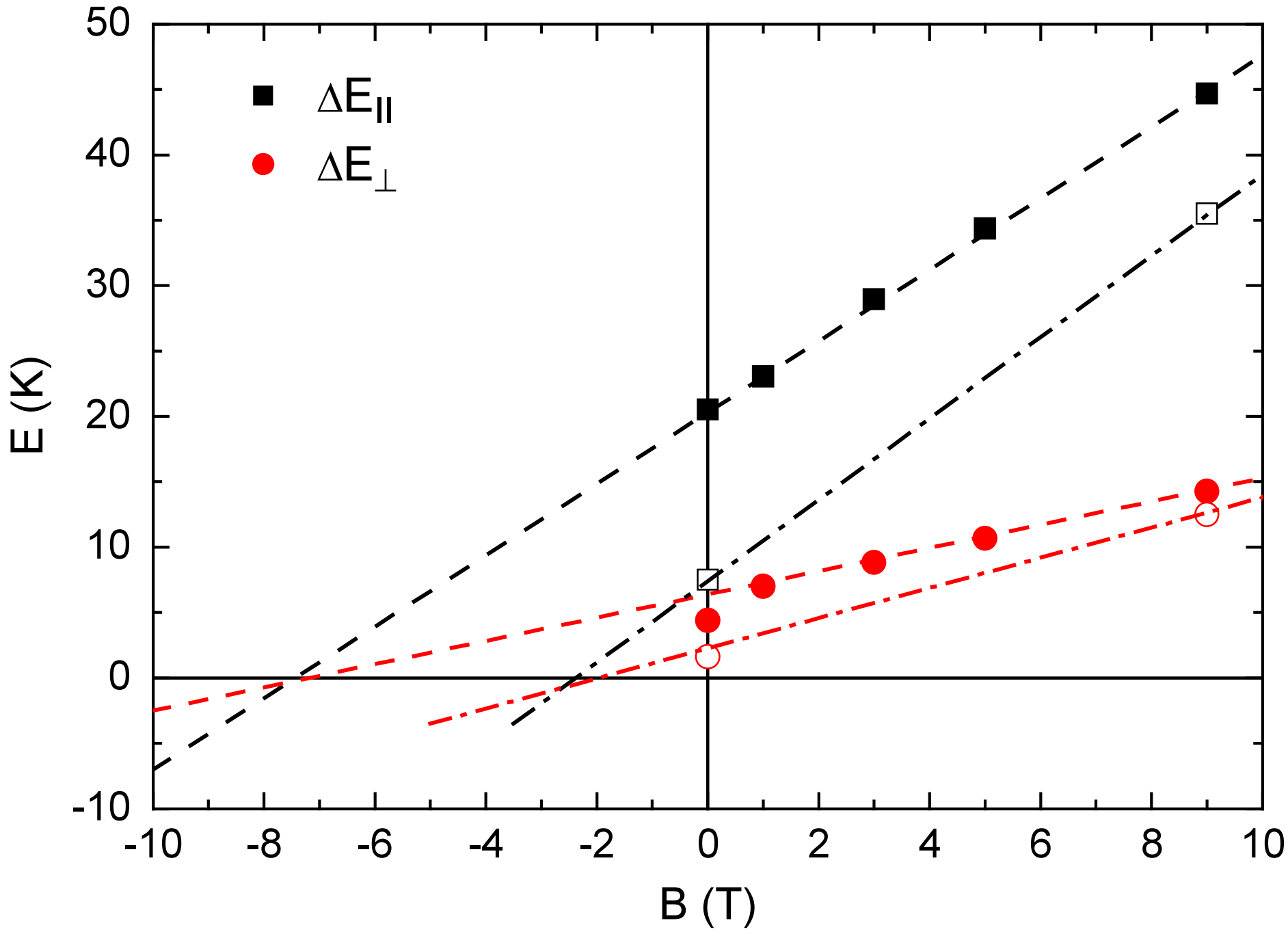

Fig.8. (Color online) Zeeman splitting ( $\Delta E_{\|}$ and $\Delta E_{\perp}$ ) of the $Pr^{4+}$ ground doublet for $Pr_{0.5}Ca_{0.5}CoO_3$ (full symbols) and $Pr_{0.63}Y_{0.07}Ca_{0.3}CoO_3$ (open symbols), determined from the fit of Schottky peaks in Fig.7.

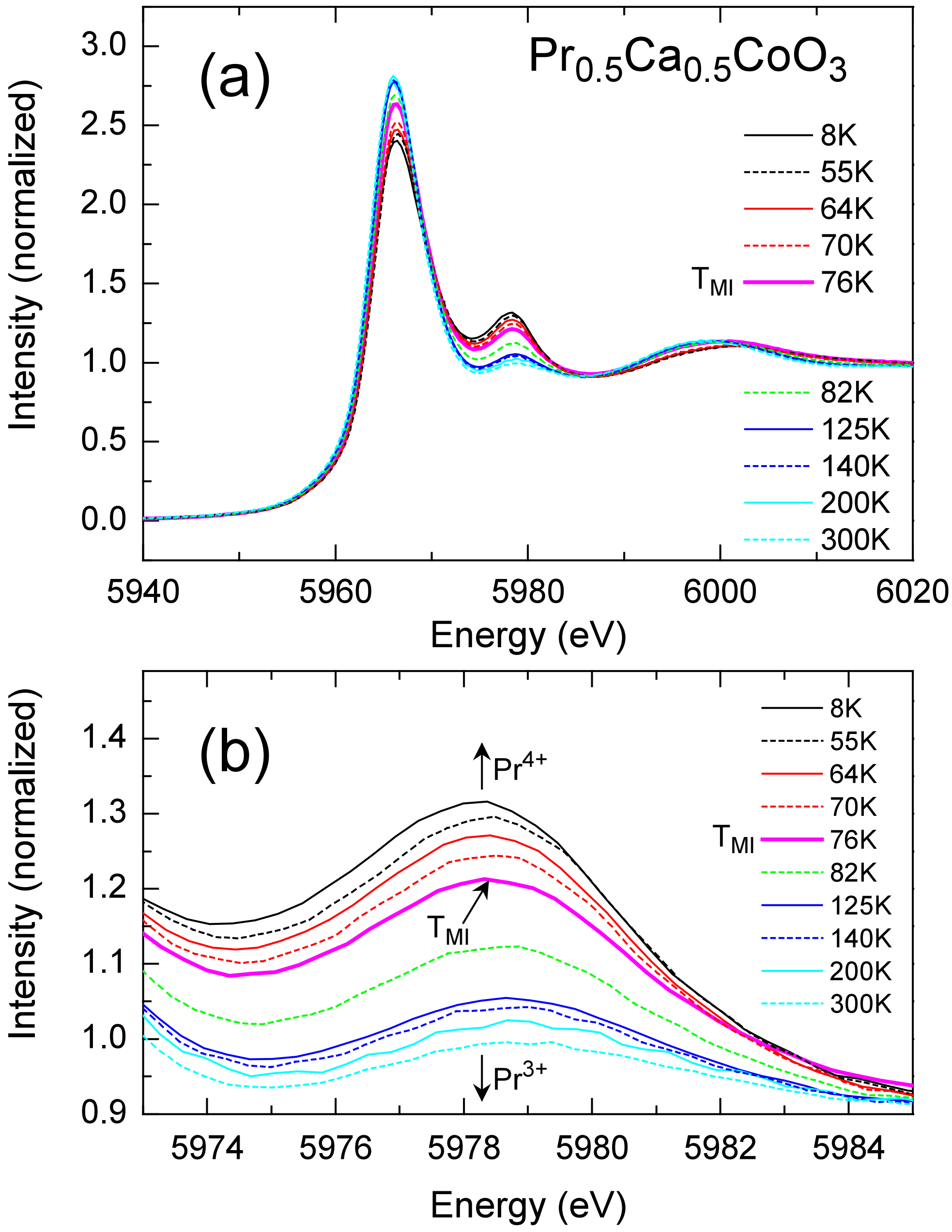

Fig.9. (Color online) The temperature dependence of the normalized XANES spectra at the Pr $L_3$-edge for $Pr_{0.5}Ca_{0.5}CoO_3$ sample. The lower panel shows the magnification of the spectra related to $Pr^{4+}$ ions.

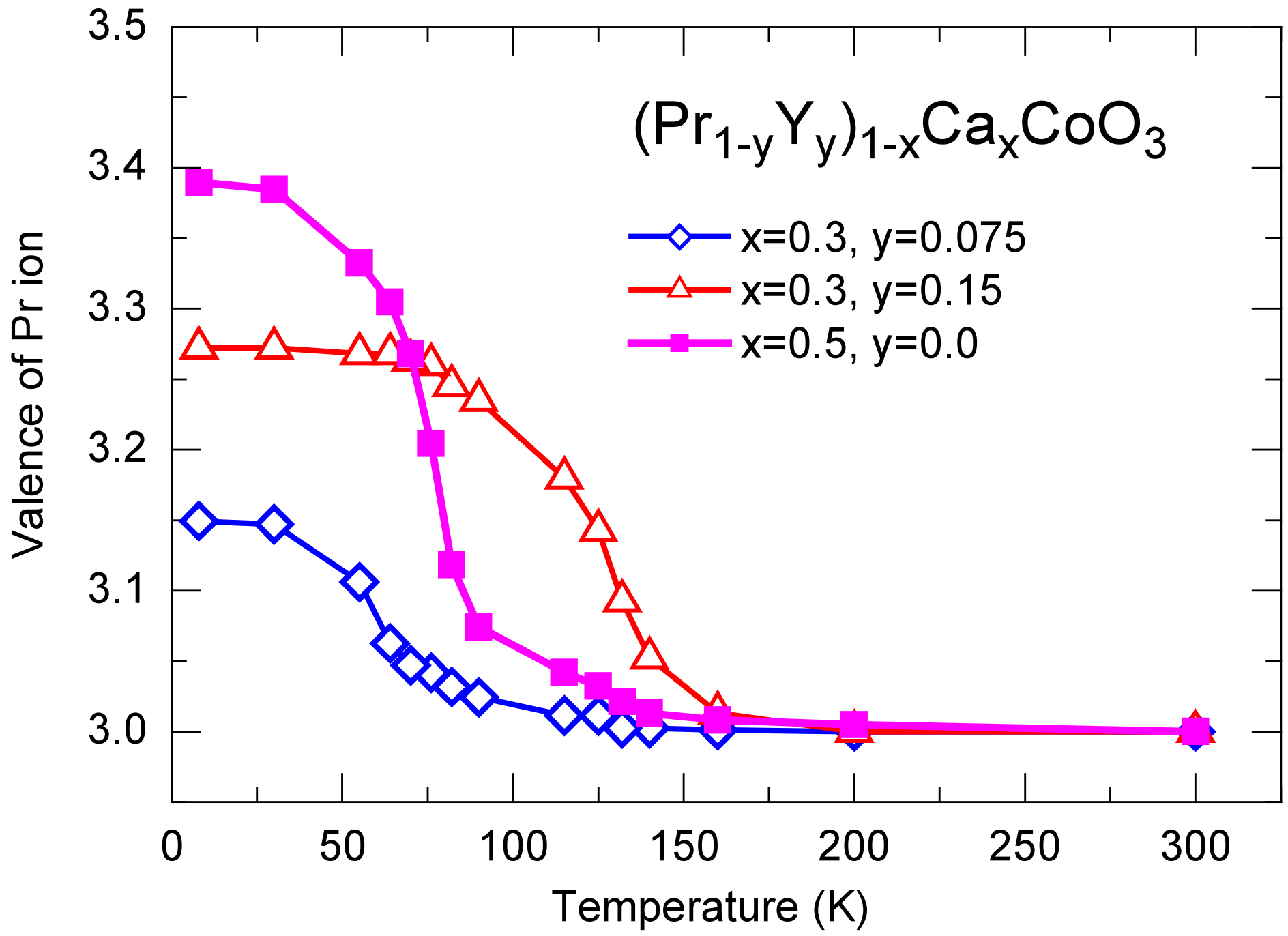

Fig.10. (Color online) The temperature dependence of the valence of Pr ions in the samples deduced from the ratio of the Lorentzian peak intensity associated with the $Pr^{3+}$ and $Pr^{4+}$ states, which fit the Pr $L_3$- edge XANES spectra. The ambiguity of the estimated valence values is ± 0:03, resuting from the arbitrariness of the parameters in the arctangent and three fitted Lorentzian peaks. For more detail we refer to the Ref.. 12.

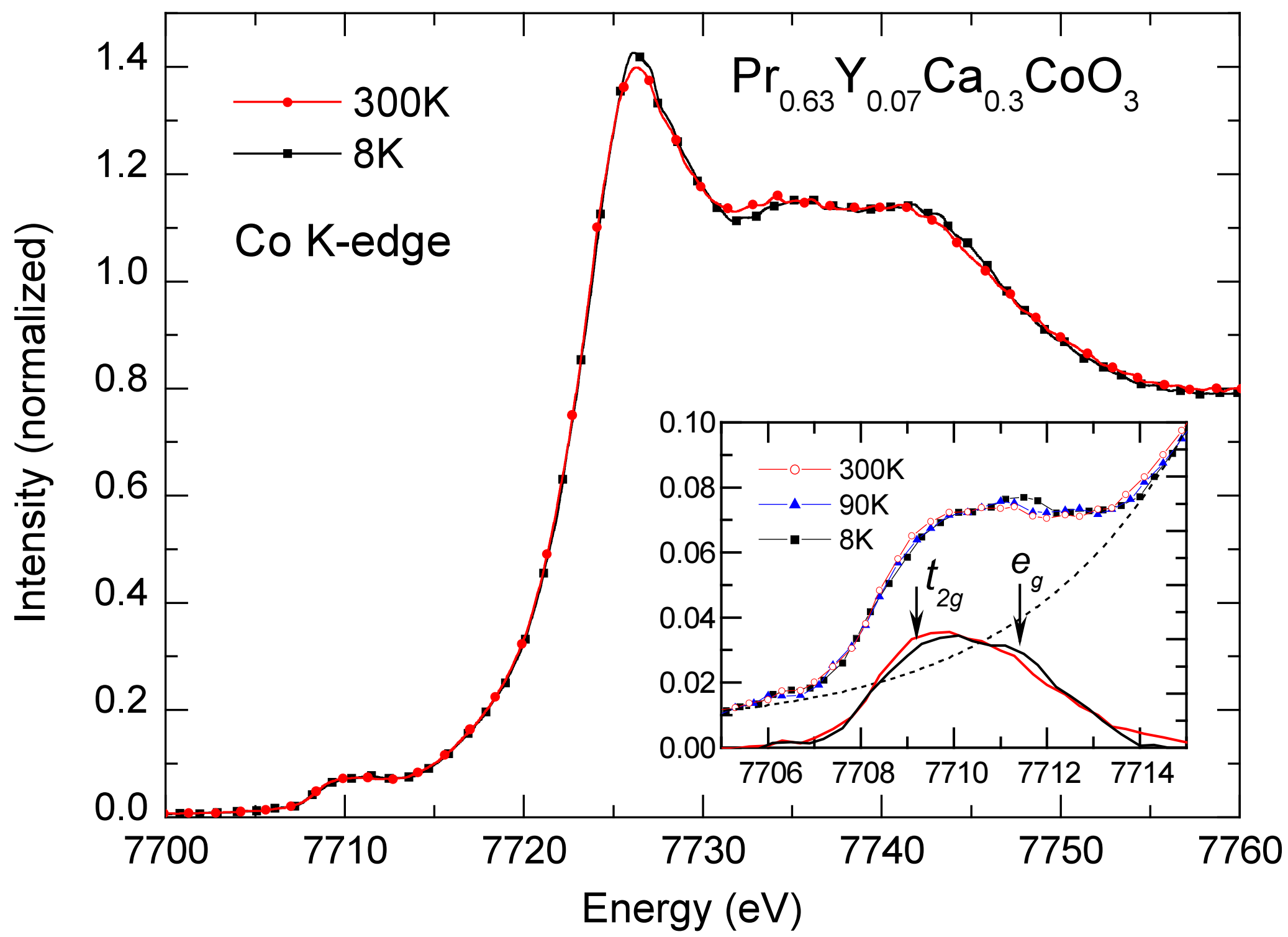

Fig.11. (Color online) The temperature dependence of the normalized XANES spectra at the Co *K*-edge for the $Pr_{0.63}Y_{0.07}Ca_{0.3}CoO_3$ sample. The inset shows the magnification of the pre-edge peak including the data after subtraction of arctangent background (full lines). Expected positions of $t_{2g}$ and $e_g$ components are marked by the arrows.